\documentclass[twocolumn, 10pt, aps, superscriptaddress, floatfix, showpacs, prb, citeautoscript]{revtex4-1}

\pdfoutput=1                    % For arXiv
\usepackage[utf8]{inputenc}
\usepackage[T1]{fontenc}
\usepackage{graphicx}
\usepackage{amsmath}
\usepackage{amssymb}
\usepackage{bm}
\usepackage{xcolor}
\usepackage[colorlinks, citecolor={blue!50!black}, urlcolor={blue!50!black}, linkcolor={red!50!black},linktoc=all]{hyperref}
\usepackage{bookmark}
\usepackage{tabularx}
\usepackage{mathtools}
\usepackage{microtype}
\usepackage[load=physical, load=abbr]{siunitx}

\DeclarePairedDelimiter\abs{\lvert}{\rvert}%
\DeclarePairedDelimiter\norm{\lVert}{\rVert}%
\DeclareMathOperator{\imag}{Im}
\DeclareMathOperator{\real}{Re}
\DeclareMathOperator{\Tr}{Tr}
\makeatletter
\let\oldabs\abs
\def\abs{\@ifstar{\oldabs}{\oldabs*}}
\let\oldnorm\norm
\def\norm{\@ifstar{\oldnorm}{\oldnorm*}}
\makeatother

\graphicspath{{figures/}}

\newcommand{\comment}[2]{#2}

\begin{abstract}
Majorana bound states (MBS) differ from the regular zero energy Andreev bound states in their nonlocal properties, since two MBS form a single fermion.
We design strategies for detection of this nonlocality by using the phenomenon of Coulomb-mediated Majorana coupling in a setting which still retains falsifiability and does not require locally separated MBS.
Focusing on the implementation of MBS based on the quantum spin Hall effect, we also design a way to probe Majoranas without the need to open a magnetic gap in the helical edge states.
In the setup that we analyze, long range MBS coupling manifests in the $h/e$ magnetic flux periodicity of tunneling conductance and supercurrent.
While $h/e$ is also the periodicity of Aharonov-Bohm effect and persistent current, we show how to ensure its Majorana origin by verifying that switching off the charging energy restores $h/2e$ periodicity conventional for superconducting systems.
\end{abstract}

\begin{document}

\title{Detecting Majorana nonlocality using strongly coupled Majorana bound states}
\author{S. Rubbert}
\affiliation{Kavli Institute of Nanoscience, Delft University of Technology,
  P.O. Box 4056, 2600 GA Delft, The Netherlands}
\author{A. R. Akhmerov}
\affiliation{Kavli Institute of Nanoscience, Delft University of Technology,
  P.O. Box 4056, 2600 GA Delft, The Netherlands}
\date{\today}


\maketitle

\section{Introduction}

\comment{Majoranas are a subject of search, but detecting their local properties is insufficient.}
The ability to create, detect, and manipulate Majorana bound states (MBS) is one of the current research goals of condensed matter physics.
MBS are the simplest non-Abelian anyons, and a potential building block of a noise-tolerant quantum computer\cite{beenakker_search_2013, alicea_new_2012, leijnse_introduction_2012}.
The experiments so far focus on identifying local properties of MBS, such as the zero bias peak in conductance\cite{mourik_signatures_2012}, the $4\pi$-periodic Josephson effect\cite{fu_josephson_2009,rokhinson_fractional_2012}, or the local maximum in the zero energy density of states\cite{nadj-perge_observation_2014}.
Observing the local signatures of MBS cleanly is an important milestone, but it has its limitations since known local signatures of MBS can be mimicked by regular Andreev bound states subjected to sufficient fine-tuning.
For instance, a topologically protected level crossing responsible for the $4 \pi$-periodic Josephson effect can be indistinguishable from an unprotected avoided level crossing\cite{sau_possibility_2012}, and a zero bias peak may have non-topological origins\cite{lee_zero-bias_2012,pikulin_zero-voltage_2012}.

\comment{Braiding is good but hard.}
Therefore an unambiguous detection of Majorana fermions requires detecting their non-local properties in a falsifiable manner.
Braiding statistics of MBS can serve as one such experiment, but even a minimal braiding setup\cite{hyart_flux-controlled_2013} requires time domain manipulation of a complicated superconducting circuit hosting six MBS, or of a large array of gate voltages\cite{alicea_non-abelian_2011}.

\comment{Another nonlocal phenomenon is the Majorana-Coulomb coupling, initially reported as teleportation.}
Another consequence of the nonlocal nature of MBS is their transport property called electron teleportation~\cite{fu_electron_2010}, discovered by L.~Fu.
It occurs in superconducting islands hosting MBS and having a finite charging energy.
If there are leads coupled to the MBS, Majorana teleportation provides coherent transport of single fermionic excitations between the leads.
The direct signatures of electron teleportation include the period doubling of a Fabry-Perot interferometer\cite{fu_electron_2010,sau_proposal_2015} and the periodicity change of the ground state energy of a ring made out of a topological superconductor\cite{van_heck_coulomb_2011}.
More advanced consequences of electron teleportation are the appearance of a high symmetry Kondo problem in multi-lead scattering off an island hosting MBS\cite{beri_topological_2012} and exotic many-body phases of a network of such islands.

\comment{We propose a setup that measures the simplest signature of teleportation.}
A simple physical interpretation of the electron teleportation is the appearance of an extra term in the Hamiltonian proportional to $i^{n/2}\prod_0^n \gamma_i$ in the presence of charging energy.
\footnote{The original publication of L. Fu, Ref.~[\onlinecite{fu_electron_2010}] uses a gauge choice that contains unphysical degrees of freedom and a fermion parity constraint, and therefore does not contain the MBS coupling term explicitly.
We use a physically equivalent gauge of Ref.~\onlinecite{van_heck_coulomb_2011}.}
In other words, the charging energy couples all the MBS $\gamma_i$ belonging to the island.
If there are only two MBS present, this coupling becomes identical to a direct overlap of low energy quasiparticle wave functions in a superconductor due to finite size effects.
In other words, charging energy coherently transports a single fermion from one MBS to another.
Since it does not require a direct wave function overlap, it is non-local.
A falsifiable detection of this non-local coupling therefore requires verification that it is coherent, that it is single fermion transport, and that it is not arising due to an actual wave function overlap.

The aim of our work is to present and analyze a setup that allows one to detect this coupling while not having any unnecessary ingredients.
Our proposed setup has an additional counter-intuitive benefit of not requiring creation of decoupled MBS, unlike required in the previous proposals\cite{fu_electron_2010,sau_proposal_2015}.
This makes our setup perfectly suited for quantum spin Hall insulator (QSHE)-superconductor hybrid structure \cite{fu_josephson_2009}, where isolation of MBS requires creation of magnetic tunnel barriers and remains an open experimental challenge.
In addition our setup allows one to distinguish the electron teleportation from local coupling through the superconductor therefore providing the falsifiability of the effect.

\section{The setup}
\subsection{System layout and qualitative arguments}
\label{SystemSec}

\comment{Let's walk through the setup requirements.}
We begin by considering each requirement for detection of the non-local coupling and arguing how to achieve it in the simplest fashion.
Once again: we aim to design a setup that has to detect coherent transport of single fermions through a topological superconductor.
Additionally it has to ensure that the origin of this transport is not due to quasiparticle current caused by a normal conduction channel.

\comment{Coherent transport is best checked by an interferometer.}
Coherence of quasiparticle transport is most directly checked by a two-path interferometer.
In order to test the electron teleportation, one arm must include the topological superconductor hosting MBS, while the other reference arm should be a normal region.
The coherence of quasiparticle transport through such an interferometer manifests in periodic dependence of observed properties on the magnetic flux threaded through it.

\comment{Aharonov-Bohm and single quasiparticle transport}
The charge of the interfering particle manifests in the flux periodicity of the interferometer's conductance and spectrum. Therefore in the presence of a conduction channel for single fermionic quasiparticles we expect an $h/e$ periodicity of the observed signal, or Aharonov-Bohm effect.
This allows us to distinguish fermion transport from Cooper pair transport flux dependence with period  $h/2e$ that corresponds to Josephson effect \cite{buttiker_flux_1986, pientka_signatures_2013, crepin_flux_2016}.

\comment{The interferometer should be open to the outside, but the Coulomb blockade should still be present.}
The $4\pi$-periodic Josephson effect arising from a fermion parity anomaly~\cite{fu_josephson_2009} may obscure the non-local coupling by creating a signal with the same periodicity.
To suppress this effect, the interferometer must be coupled to a normal metallic reservoir draining out of equilibrium fermionic excitations.
On the other hand, the coupling of the superconductor to an external reservoir cannot have high transparency, since then a low $RC$-time suppresses the Coulomb blockade and the non-local coupling.
These two requirements are satisfied if a tunnel junction is present between the superconductor and the normal reservoir. 
In the setups of Refs.~\onlinecite{fu_electron_2010,sau_proposal_2015} the tunnel barrier separates the two interferometer arms and suppresses the coupling strength $E_\textrm{M}$ through the reference arm.
Locating the tunnel barrier directly between the normal interferometer arm and the metallic lead avoids the coupling strength suppression and simplifies the setup \footnote{Two possible strategies to realize the normal lead weakly coupled to a quantum spin Hall edge are either creating a dielectric-normal metal tunnel junction, or a gate-tunable narrow constriction in the quantum spin Hall edge itself.}.

\comment{We rule out normal transport by turning off the Coulomb energy.}
The final requirement our setup should satisfy is the need to rule out the conventional quasiparticle transport through the nontrivial part of the interferometer.
Since the quasiparticle transport appears also without Coulomb energy, suppressing the latter and observing disappearance of the $h/e$ interference signal allows one to conclude that the interference is of non-local origin.
We propose to use a standard technique~\cite{koch_charge_2007} to controllably suppress the Coulomb energy $E_C$ by adding a flux- or gate-tunable \cite{larsen_semiconductor-nanowire-based_2015} Josephson coupling $E_\textrm{J}$ between the nontrivial interferometer arm and a superconducting reservoir.
This leads to a renormalization~\cite{koch_charge_2007} of the effective charging energy $\tilde{E}_C \propto \exp(-\sqrt{8E_\textrm{J}/E_C})$ when $ E_\textrm{J} \gg E_C$.

\begin{figure}[tbh]
\includegraphics[width=\linewidth]{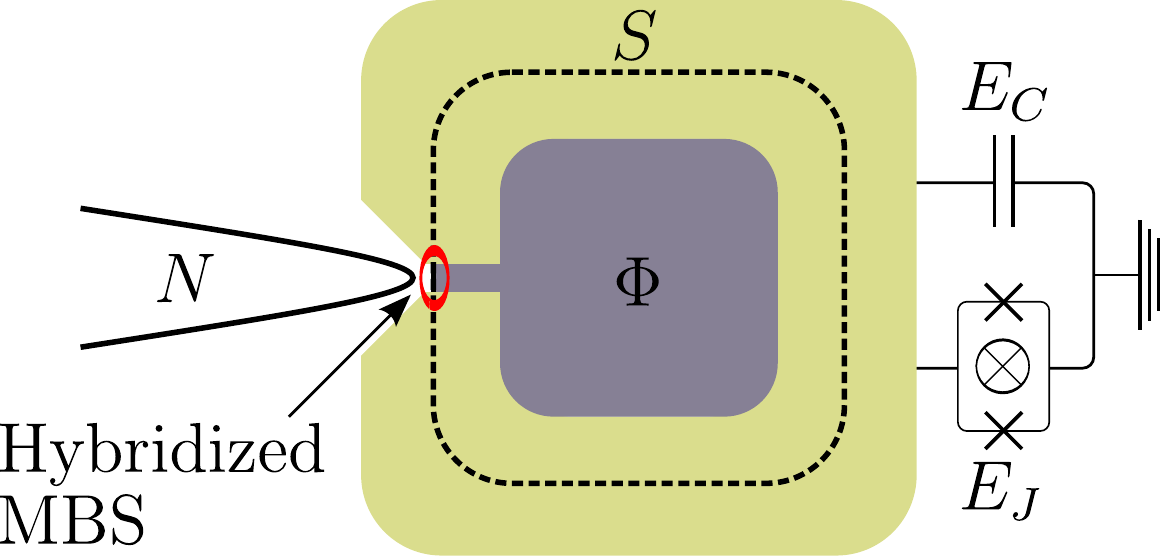}
\caption{
\label{SketchSystemFig}
Setup consists of a QSHE insulator (dark gray) with its edge (dashed line) partially covered by a superconducting ring.
The proximity-induced gap in the QSHE edge forms two hybridized MBS at the part of the edge not covered by the superconductor. 
A tunable Josephson junction couples the superconducting ring to the superconducting lead.
Finally, the normal lead weakly couples to the quantum spin Hall edge in the junction region.}
\end{figure}

\comment{Bringing everything together we get a decent setup with easy to measure observables.}
We arrive at the setup shown in Fig.~\ref{SketchSystemFig}, that consists of an interferometer coupled to a normal lead by a tunnel junction and a superconducting lead by a tunable Josephson junction, for example a dc-SQUID.
Every element in this system may only be replaced and not removed because all of them have a separate role in detection of nonlocal signatures of MBS.
The effective low energy Hamiltonian of this system is
\begin{align}
\label{excitationEq}
H_\textrm{eff} &= i\gamma_1\gamma_2\left[E_\textrm{M}\cos \left(\pi\Phi /\Phi_0\right) + \tilde{E}_C \cos \left(\pi n_I \right)\right] \\
             &\equiv i\gamma_1\gamma_2 \Delta E,\nonumber
\end{align}
with $\gamma_1,\gamma_2$ the Majorana operators, $n_I$ the induced charge of the interferometer, $\Phi$ the flux through it, and $\Phi_0 = h/2e$ the superconducting flux quantum.
When $\tilde{E}_C$ is finite, the spectrum of this Hamiltonian is $h/e$-periodic in $\Phi$, but it becomes $h/2e$-periodic when $\tilde{E}_C$ is suppressed by increasing $E_\textrm{J}$.
A corresponding Hamiltonian of a trivial Josephson junction containing a single Andreev bound state has a form $H = [E_J(\Phi)+E_C]a^\dagger a$, where $E_J$ is a $h/2e$-periodic function of $\Phi$ and $a$ is the annihilation operator of the Andreev bound state, so that its periodicity is always constant.
Quasiparticles tunneling through the superconductor give rise to a term $it_{\textrm{SC}}\gamma_1\gamma_2$ with $t_{\textrm{SC}}$ the tunneling amplitude, and keep the spectrum $h/e$-periodic regardless of $\tilde{E}_C$.
As we will show in more detail, measuring either the supercurrent circulating in the interferometer or the conductance between the normal and the superconducting leads as a function of flux reveals the periodicity of the spectrum and provides an observable signature of the nonlocal properties of Majorana fermions.

\subsection{Effective Hamiltonian}
\label{HamiltonianSec}
\comment{Cooper Box+Lead Hamiltonian with explanation of variables/operators}
The effective Hamiltonian of the Coulomb Majorana interferometer of Fig.~\ref{SketchSystemFig} is
\begin{equation}
H = H_{\textrm{CPB}} + \sum_{k,\sigma}\varepsilon(k, \sigma)c_{k,\sigma}^{\dagger}c_{k,\sigma} + H_{\textrm{c}}\label{FirstHamiltonianEq}.
\end{equation}
Here $H_{\textrm{CPB}}$ is the Cooper pair box Hamiltonian:
\begin{align}
H_{\text{CPB}} &= E_C \left(-2i\partial_\phi + n_I + p/2 \right)^{2} - E_\textrm{J}\cos\phi \label{OnsiteHamiltonianEq}\\
  &- E_{\textrm{M}} p\cos\left(\pi \Phi/\Phi_0 \right),\nonumber
\end{align}
and $H_\textrm{c}$ is the coupling Hamiltonian between the Cooper pair box and the normal lead
\begin{equation}
H_\textrm{c} = \sum_{k,\sigma} \bigg[c_{k,\sigma} e^{i(1-p)\phi/2} (t_{\sigma, 1}\gamma_1 +t_{\sigma,2}\gamma_2)+h.c. \bigg].\label{couplingham}
\end{equation}
Here $\phi$ is the superconducting phase of the island, and $p=i\gamma_1 \gamma_2$ is the fermion parity of the interferometer.
Finally, the tunnel coupling between the lead modes and the MBS is $t_{\sigma, i}$, and it may depend on $\phi$.

\comment{Break down to first quantization/single particle physics}
Rewriting the Hamiltonian in the eigenbasis of the Cooper-pair box yields:
\begin{widetext}
  \begin{subequations}
    \label{eq:eigenbasis_hamiltonian}
    \begin{align}
      H_{\textrm{CPB}} &=  E_\Phi(p, b^\dagger b), \\
      H_{\textrm{c}} &= \sum_{k,\sigma, n} \left[c_{k,\sigma} \left(t_{\sigma, 1}\gamma_1 + t_{\sigma, 2}\gamma_2 \right)
                       \left[\xi_n^+(p, b^\dagger b)(b^\dagger)^n + \xi_n^-(p, b^\dagger b)b^n\right] + \textrm{h.c.} \right],
                       \label{SecondHamiltonianEq}\\
      \xi_n^{\pm}(p, m) &\equiv \langle m\pm n, -p | e^{i(1-p)\phi/2 } |m, p\rangle.
    \end{align}
  \end{subequations}
\end{widetext}
Here we introduced the eigenenergies of the Cooper-pair box $E_\Phi(i\gamma_1\gamma_2, b^\dagger b)$ and the ladder operators of the Cooper-pair box $b$ and $b^\dagger$.
An electron/hole tunneling into the superconducting ring can create excitations in the Cooper-pair box.
In the Eq.~\eqref{eq:eigenbasis_hamiltonian} this is expressed by the transition amplitudes $\xi$  between the states $|n, p\rangle$, with $n$ the number of Cooper pair box excitations, and $p$ its fermion parity.
In the following we calculate $\xi$ and $E(i\gamma_1\gamma_2, b^\dagger b)$ numerically (for details of our numerical calculations see the Supplemental Material available with the manuscript).

\section{Readout}

\subsection{Zero bias conductance}
\label{ConductanceCalculationSec}
\comment{Hamiltonian for a single excitation}
The observable steady-state properties in this system, such as conductance, in general have the same flux periodicity as the spectrum, and therefore should exhibit signatures of the non-local coupling.
However, evaluating conductance at an arbitrary bias is an involved task and to simplify the calculation we focus on the zero bias.
Since the quasiparticle lifetime in the interferometer is bounded from above by the inverse coupling to the lead, simultaneous tunneling events of multiple quasiparticles are suppressed at voltages $eV \ll |t|$. 
Therefore in this regime we may project the Hamiltonian onto the Hilbert space of a single fermionic excitation in order to simplify the problem. 
The basis states of the single fermion Hilbert space are:
\begin{subequations}
\label{eq:single_particle_basis}
\begin{align}
|k, \sigma, e \rangle = c^\dagger_{k, \sigma} |\text{gs}_{\text{lead}}\rangle \otimes |\text{gs}_{\text{ring}} \rangle , \\
|k, \sigma, h \rangle = c_{k, \sigma} |\text{gs}_{\text{lead}}\rangle \otimes |\text{gs}_{\text{ring}} \rangle , \\
|n \rangle = (\gamma_1 + p_{\textbf{gs}} i\gamma_2) (b^\dagger)^n  |\text{gs}_{\text{lead}}\rangle \otimes |\text{gs}_{\text{ring}} \rangle.
\end{align}
\end{subequations}
Here $ |\text{gs}_{\text{lead}}\rangle $ and $ |\text{gs}_{\text{ring}} \rangle$ are the ground states of the lead and the superconducting ring and $p_{\text{gs}} = \langle \text{gs}_{\text{ring}}| p | \text{gs}_{\text{ring}} \rangle$.
The indices $e$ and $h$ correspond to the electron and hole excitations.
Projecting the Hamiltonian of Eq.~\eqref{eq:eigenbasis_hamiltonian} on the basis states of Eq.~\eqref{eq:single_particle_basis} we obtain:
\begin{align}
H_{\text{sqp}} = & \sum_{k, \sigma, \tau} |k, \sigma, \tau\rangle \tau_z \varepsilon(k, \sigma)  \langle k, \sigma, \tau |  \nonumber \\
+ & \sum_n |n \rangle E_\Phi(-p_{\text{gs}}, n) \langle n| \\
+ & \sum_{k, \sigma, \tau, n} \left[ | n \rangle \chi_n(p_{\text{gs}}, \tau) \langle k, \sigma, \tau |  + h.c.\right], \nonumber
\label{first_quantized_Hamiltonian}
\end{align}
with
\begin{align}
\chi_n(p, \tau) &= \langle n | H_{\textrm{c}}  | k, \sigma, \tau \rangle.
\end{align}
Because of the doubling of degrees of freedom, $\chi$ also depends on the particle-hole index $\tau$, even though the previously defined $\xi$ does not.

\comment{scattering matrix}
We use the Mahaux-Weidenm\"uller formula to calculate the scattering matrix:
\begin{eqnarray}
S=\frac{1 + i\pi W^{\dagger}(\sum_n |n \rangle  E_\Phi(-p_{\text{gs}}, n)  \langle n|-E)^{-1}W}{1 - i\pi W^{\dagger}(\sum_n |n \rangle E_\Phi(-p_{\text{gs}}, n) \langle n|-E)^{-1}W}.
\label{Mahaux-WeidenmüllerSquareEq}
\end{eqnarray}
Here $E$ is the quasiparticle energy, and $W$ is the coupling to the leads
\begin{eqnarray}
W = \sqrt{\rho} \sum_{\sigma, \tau, n} | n \rangle \chi_n(p_{\text{gs}}, \tau) \langle k_E, \sigma, \tau |,
\end{eqnarray}
with $\rho = (d \varepsilon /d k)^{-1}$ the density of states in the lead and $k_E$ the momentum of excitations at energy $E$.

\begin{figure}[tb]
\includegraphics[width=\linewidth]{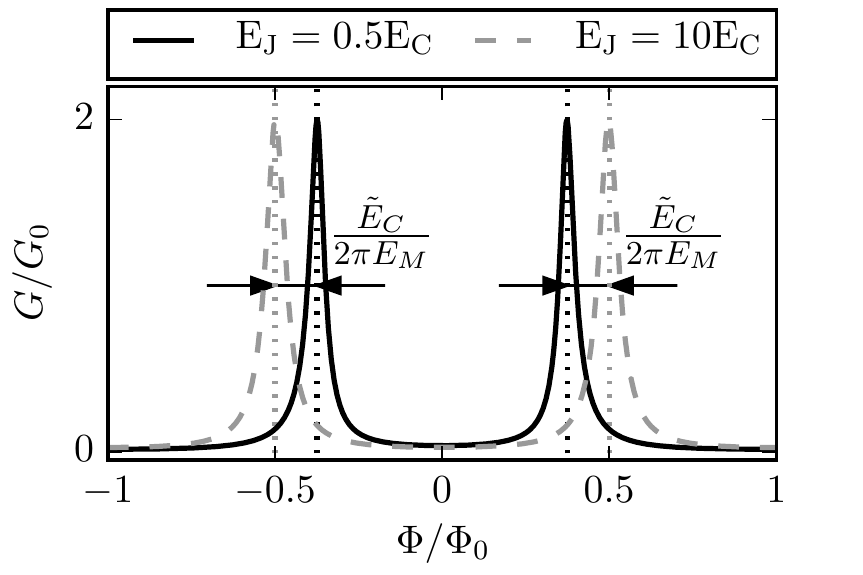}
\caption{
\label{ConductanceFig}
Conductance of the interferometer of Fig.~\ref{SketchSystemFig} as a function of magnetic flux  through the superconducting ring.
It has an $h/e$ periodicity if the effective charging energy $\tilde{E}_C$ is not suppressed.
This numerical calculation follows Appendix~\ref{Appendix A}, where more than one excited state of the Cooper-pair box was taken into account.}
\end{figure}

\comment{Conductance}
The differential conductance of the device is $G = 2G_0 \Vert S_{he}\Vert^2$, with $G_0 = e^2/h$ the conductance quantum.
If the tunneling amplitude is much smaller than the level spacing in the ring, $H_{\text{S}}$ is well approximated by truncating it to the two lowest energy states with opposite fermion parity.
It yields the conductance of a resonant Andreev level
\begin{equation}
G = \frac{2G_0}{1+\Delta E^{2}/(\pi^2 \lVert W\rVert^4)},
\end{equation}
with $\Delta E$ the splitting between the Majorana states, given by Eq.~\eqref{excitationEq}.
The resonant peaks appear when $\Delta E = 0$, and therefore they have $h/2e$ periodicity in absence of the nonlocal coupling $\tilde{E}_C$ that changes into $h/e$ when $\tilde{E}_C \gtrsim \lVert W\rVert$.
Andreev conductance calculated using the full excitation spectrum of the ring (see Appendix~\ref{Appendix A}) is shown in Fig.~\ref{ConductanceFig}, and it qualitatively agrees with the behavior of the two-level system.
Since the flux dependence of the tunneling amplitudes has to have a period of $h/2e$, it does not impact our result.

\subsection{Supercurrent}

\comment{What other measurement}
Supercurrent carried by the interferometer in its ground state is also sensitive to the $h/e$ periodicity of the Hamiltonian.
It can be measured using SQUID magnetometry\cite{sochnikov_nonsinusoidal_2015}, and is thus an alternative pathway to observe the nonlocal coupling of Majoranas in the same interferometer.
The current with $h/e$ periodicity in the interferometer is an equilibrium phenomenon, and therefore different from the $4\pi$-periodic Josephson effect, which is a non-equilibrium effect appearing due to a fermion parity anomaly.
Since the coupling to the normal lead breaks the fermion parity conservation, it also suppresses the $4\pi$-periodic Josephson effect in the interferometer.

\begin{figure}
\includegraphics[width=\linewidth]{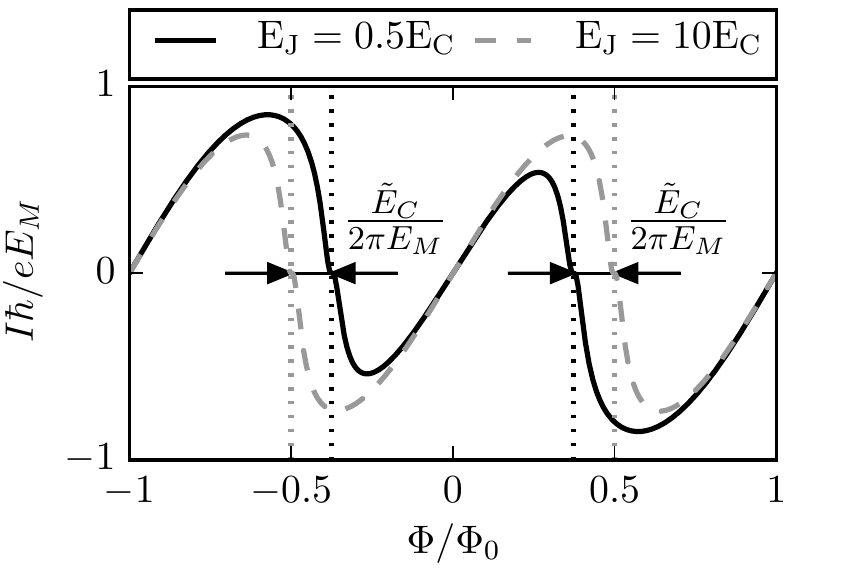}
\caption{
\label{CurrentResponseFig}
Supercurrent response of the interferometer of Fig.~\ref{SketchSystemFig} to a magnetic flux  through the superconducting ring.
The current-flux relationship has a period of $h/e$ if the effective charging energy is not negligible (black solid line).
The Josephson $h/2e$ periodicity is restored when $\tilde{E}_C$ is suppressed by a large $E_\textrm{J}$.
The supercurrent vanishes near the level crossing of the even and odd parity ring states; the low energy spectrum is symmetric around that flux.
}
\end{figure}

\comment{Outline calculation}
We calculate the supercurrent in the ring using the definition
\begin{align}
I=\frac{\partial E_\textrm{gs}}{\partial\Phi},
\end{align}
with $E_\textrm{gs}$ the ground state energy of the interferometer including the lead.
We obtain $E_\textrm{gs}$ by integrating the density of states
\begin{align}
\frac{\partial n}{\partial E}=\frac{1}{2\pi}\imag\Tr\frac{\partial S_{\alpha\beta}^{\dagger}}{\partial E}S_{\alpha\beta}
\label{density_of_statesEq}
\end{align}
over negative quasiparticle energies (see Appendix~\ref{Appendix B} for details).
The resulting current-flux relationship is shown in Fig.~\ref{CurrentResponseFig}, and in agreement with our expectations we observe that a finite effective capacitive energy makes supercurrent $h/e$-periodic.

\subsection{Parameter value estimation}

The Majorana coupling in a short junction is comparable to the induced superconducting gap, $E_\textrm{M} \approx \Delta$ \cite{2009PhRvB..79p1408F}.
Maximizing $E_\textrm{M}$ is unfavorable for the observation of nonlocal coupling since the magnitude of the $h/e$-periodic component is proportional to $\tilde{E}_C/E_\textrm{M}$.
This argument together with the high availability of Al make it the optimal superconductor for observing the nonlocal coupling, and hence we use $E_\textrm{M} \approx \Delta_{\textrm{Al}} \approx \SI{0.1}{\meV}$.

We assume that the capacitance is dominated by the coupling between the superconductor and the back-gate required to tune the quantum spin Hall device into the insulating regime.
If the superconducting ring has a circumference $L=\SI{3}{\um}$, width $w=\SI{0.1}{\um}$, the distance to the gate is $d=\SI{0.1}{\um}$, and the gate dielectric has $\varepsilon_r = 10$, then the capacitance $C = \varepsilon_0 \varepsilon_r Lw/ d \approx \SI{1.8}{\femto\farad}$, or $E_C \approx \SI{0.1}{\meV}$.
The bare Coulomb energy is comparable to $E_\textrm{M}$, and therefore the Josepshon energy should change within a range between $ E^{\text{Max}}_J \gtrsim 10 E_C $ and $ E^{\text{Min}}_J \lesssim E_C$.

Finally, the coupling strength of the normal lead to the MBS needs to be smaller than the energy scales $E_\textrm{M}$ and $\tilde{E}_C$, since otherwise the ground and excited states are overlapping due to level broadening.

\section{Summary}
Due to the experimental progress towards the controllable creation of MBS, the planning of next steps in coherent control of MBS becomes a timely and relevant question.
The currently existing proposals include braiding\cite{hyart_flux-controlled_2013}, a simpler non-topological qubit rotation\cite{heck_minimal_2015}, or a Bell inequality violation\cite{clarke_practical_2015}.
We have developed an alternative measurement aiming to probe the nonlocal properties of MBS focusing on simplicity and falsifiability.
While being applicable to any implementation of MBS, our proposal has an additional advantage in quantum spin Hall devices, because it does not require spatial separation of MBS or inducing a magnetic gap in the edge states.

Our proposed setup is a Coulomb Majorana interferometer that measures a known phenomenon of Majorana teleportation through appearance of Aharonov-Bohm periodicity of conductance or supercurrent.
According to our estimates such an interferometer can be made using existing fabrication techniques and provide a sufficiently strong nonlocal signal.

\acknowledgments

We thank L.~P.~Kouwenhoven, S.~Nadj-Perge, and A.~Yacoby for discussions and useful comments. This research was supported by the Foundation for Fundamental Research on Matter (FOM), the Netherlands Organization for Scientific Research (NWO/OCW), and an ERC Starting Grant.

\bibliographystyle{apsrev4-1}
\bibliography{Bibliography.bib}

\appendix
\section{Multiple Cooper-pair box states}
\label{Appendix A}

\comment{What do we do?}
We relax the restriction of Sec.~\ref{ConductanceCalculationSec} that only takes one excited TSC ring state into account by considering the full spectrum of the Cooper-pair box.
\comment{Calculate}
This yields
\begin{multline}
\sum_{\sigma,\sigma'}|S_{e,h,\sigma,\sigma'}|^{2}= \\
\sum_{\sigma,\sigma'} \left|\sum_{n,n'} \frac{4 W_{n,\sigma, e}W^*_{n, \sigma', h}W_{n', \sigma, e}W^*_{n', \sigma', h}}{ \left( \frac{i}{\pi}H_n+\lVert W_{n} \rVert^{2} \right)
\left( \frac{-i}{\pi}H_{n'}+\lVert W_{n'}\rVert^{2} \right)}\right|,
\end{multline}
with $H_n = E_{\Phi}(-p_{\text{gs}, n})$ and $W_n$ is the $n$-th row of $W$.
Since the relative phases between $W_{n, \sigma, \tau}$ and $W_{n, \sigma', \tau'}$ do not  depend on $n$, we interchange the absolute value and the sum, arriving at:
\begin{widetext}
\begin{align}
\sum_{\sigma,\sigma'}|S_{e,h,\sigma,\sigma'}|^{2}
= & \sum_{n,n'} \frac{4\sum_{\sigma,\sigma'}|W_{n,\sigma, e}W_{n,\sigma', h}W_{n',\sigma, e}W_{n',\sigma',h}|}{ \sqrt{\left(\frac{1}{\pi^2}E^2+\lVert W_{n}\rVert^{4} \right) \left(\frac{1}{\pi^2}H_{n'}^2+\lVert W_{n'}\rVert^{4} \right)}}
= \sum_{n,n'} \frac{\lVert W_n\rVert^2 \lVert W_{n'}\rVert^2}{ \sqrt{\left(\frac{1}{\pi^2}H_n^2+\lVert W_{n}\rVert^{4} \right) \left(\frac{1}{\pi^2}H_{n'}^2+\lVert W_{n'}\rVert^{4} \right)}}.
\end{align}
\end{widetext}
\comment{discussion}
Each of the excited states of the ring yields a Lorentzian contribution to the conductivity.
In addition there are interference contributions for $n \neq n'$ that are suppressed if $|H_n-H_{n'}| \gg \lVert W_n\rVert^2$ or $|H_n-H_{n'}| \gg \lVert W_{n'}\rVert^2$.

\section{Magnetic response}
\label{Appendix B}

In this section we calculate the interferometer magnetic response, shown in Fig.~\ref{CurrentResponseFig}.
The ground state energy $E = E_0 + E_1$ has contributions $E_0$ and $E_1$ from the lowest even and odd parity states (we neglect higher energy states). We find $E_0$ and $E_1$ by calculating the local density of states in the ring using Eq.~\eqref{density_of_statesEq} and integrating over the energy of all occupied states

\begin{widetext}
\begin{align}
E_{0}=\int_{-\infty}^{0} (E+H_{0}) \real \left[ \frac{1-2\pi i\lVert W\rVert^{2} \left(H_{1}-H_{0}-E+i\pi\lVert W\rVert^{2} \right)^{-1}}{\lVert W\rVert^{2} \left(H_{1}-H_{0}-E-i\pi\lVert W\rVert^{2} \right)^{2}} \right] dE, \\
E_{1}=\int_{-\infty}^{0} (E+H_{1}) \real \left[ \frac{1-2\pi i\lVert W\rVert^{2}\left(H_{0}-H_{1}-E+i\pi\lVert W\rVert^{2} \right)^{-1}}{\lVert W\rVert^{2} \left(H_{0}-H_{1}-E-i\pi\lVert W\rVert^{2} \right)^{2}} \right] dE.
\end{align}
Here $H_0$ and $H_1$ are the energies of the ring without level broadening. The expressions are equivalent, except for interchanging $H_0$ and $H_1$.
We calculate the supercurrent using the definition $I=(2e/\hbar)\partial_{\Phi}E$, so we need to calculate
\begin{align}
\partial_{\phi}E_{0}=\partial_{\phi}\int_{-\infty}^{H_{0}-H_{1}}(E+H_{1})\left[\frac{\lVert W\rVert^{2}}{E{}^{2}+\pi^{2}\lVert W\rVert^{4}}\right]dE . 
\end{align}
Evaluating this integral and summing the contributions of both states yields
\begin{align}
\partial_{\phi}E=\frac{\lVert W\rVert^{2} \left(\partial_\Phi H_0-\partial_\Phi H_1 \right) \left( H_{0}-H_{1} \right)}{(H_{0}-H_{1})^{2}+\pi^{2}\lVert W\rVert^{4}}+ \left(\partial_\Phi H_1-\partial_\Phi H_0 \right)\frac{1}{\pi}\arctan\left(\frac{H_{0}-H_{1}}{\pi\lVert W\rVert^{2}}\right)+\frac{1}{2} \left(\partial_\Phi H_1+\partial_\Phi H_0 \right).
\end{align}
\end{widetext}
\end{document}